\documentstyle[prl,epsf,eqsecnum,aps,floats,amsmath]{revtex}  % use this for PRL mode
\begin{document}
% Change equation numbers to single arabic
\def\theequation{\arabic{equation}}%
% Rapidity Gap Definitions
\newcommand{\jjmass}{\mbox{$M_{\rm JJ}$}}
\newcommand{\rstev}{\mbox{$\rs = \T{1.8}$}}
\newcommand{\XX}{\mbox{$\, \times \,$}}
\newcommand{\AP}{\mbox{${\rm \bar{p}}$}}
\newcommand{\SU}{\mbox{$S$}}
\newcommand{\SPt}{\mbox{$<\! |S|^2 \!>$}}
\newcommand{\ET}{\mbox{$E_{T}$}}
\newcommand{\PT}{\mbox{$p_{t}$}}
\newcommand{\DP}{\mbox{$\Delta\phi$}}
\newcommand{\DR}{\mbox{$\Delta R$}}
\newcommand{\DE}{\mbox{$\Delta\eta$}}
\newcommand{\DEP}{\mbox{$\Delta\eta_{c}$}}
\newcommand{\DEC}{\mbox{$\Delta\eta_{c}$}}
\newcommand{\SP}{\mbox{$S(\DEP)$}}
\newcommand{\PH}{\mbox{$\phi$}}
\newcommand{\EA}{\mbox{$\eta$}}
\newcommand{\EAJ}{\mbox{\EA(jet)}}
\newcommand{\AEA}{\mbox{$|\eta|$}}
\newcommand{\Ge}[1]{\mbox{#1 GeV}}
\newcommand{\T}[1]{\mbox{#1 TeV}}
\newcommand{\D}[1]{\mbox{$#1^{\circ}$}}
\newcommand{\x}{\cdot}
\newcommand{\ra}{\rightarrow}
% units
\newcommand{\mb}{\mbox{mb}}
\newcommand{\nb}{\mbox{nb}}
\newcommand{\ipb}{\mbox{${\rm pb}^{-1}$}}
\newcommand{\inb}{\mbox{${\rm nb}^{-1}$}}
\newcommand{\rs}{\mbox{$\sqrt{s}$}}
\newcommand{\fdel}{\mbox{$f(\DEP)$}}
\newcommand{\fdele}{\mbox{$f(\DEP)^{exp}$}}
\newcommand{\fgap}{\mbox{$f(\DEP\! > \!3)$}}
\newcommand{\fgape}{\mbox{$f(\DEP\! > \!3)^{exp}$}}
\newcommand{\fpyt}{\mbox{$f(\DEP\!>\!2)$}}
\newcommand{\delth}{\mbox{$\DEP\! > \!3$}}
\newcommand{\uplim}{\mbox{$1.1\!\times\!10^{-2}$}}
\newcommand{\sigew}{\mbox{$\sigma_{\rm EW}$}}
\newcommand{\sigsi}{\mbox{$\sigma_{\rm singlet}$}}
\newcommand{\sigr}{\mbox{$\sigsi/\sigma$}}
\newcommand{\sigrew}{\mbox{$\sigew/\sigma$}}
\newcommand{\ncal}{\mbox{$n_{\rm cal}$}}
\newcommand{\ntrk}{\mbox{$n_{\rm trk}$}}
% greater/less than or approximately equal to 
\def\simge
{\mathrel{\rlap{\raise 0.53ex \hbox{$>$}}{\lower 0.53ex \hbox{$\sim$}}}}

\def\simle
{\mathrel{\rlap{\raise 0.4ex \hbox{$<$}}{\lower 0.72ex \hbox{$\sim$}}}}

%\input D0$ROOT:[D0TEX]D0_STYLE.TEX
% D0 Style Definitions
% Definitions of commonly used symbols
%
\def\sigtot{$\sigma_{\rm tot}$}         %sigma total
\def\sigtop{$\sigma_{t \overline{t}}$}  %sigma_ttbar
\def\pbarp{$\overline{p}p $}            %pbarp
\def\ppbar{$p\overline{p} $}            %ppbar
\def\qqbar{$q\overline{q}$}             %qqbar
\def\qbarq{$\overline{q}q$}             %qbarq
\def\ttbar{$t\overline{t}$}             %ttbar
\def\bbbar{$b\overline{b}$}             %bbbar
\def\D0{D\O}                            %D0
\def\CDF{CDF}
\def\ipb{pb$^{-1}$}                     %inverse picobarns
\def\pt{p_T}                            %pT
\def\ptg{p_T^\gamma}                    %pT_gamma
\def\et{E_T}                            %ET
\def\etg{E_T^\gamma}                    %ET_gamma
\def\htran{$H_T$}                       %HT
\def\gevcc{{\rm GeV}/c^2}               %GeV/c^2
\def\gevc{{\rm GeV}/c}                  %GeV/c
\def\gev{~\rm GeV}                       %GeV
\def\tev{~\rm TeV}                       %TeV
\def\njet{$N_{\rm jet}$}                %N_jet
\def\aplan{$\cal{A}$}                   %aplanarity
\def\lum{$\cal{L}$}                     %luminosity
\def\iso{$\cal{I}$}                     %isolation variable
\def\remu{${\cal{R}}_{e\mu}$}           %distance between e and mu in eta-phi
\def\rmu{$\Delta\cal{R}_{\mu}$}         %distance between mu and jet cone axis
\def\pbar{$\overline{p}$}               %pbar
\def\tbar{$\overline{t}$}               %tbar
\def\bbar{$\overline{b}$}               %bbar
\def\lumint{$\int {\cal{L}} dt$}        %integrated luminosity
\def\lumunits{cm$^{-2}$s$^{-1}$}        %luminosity units
\def\etal{{\sl et al.}}                 %et al. - no preceeding comma
\def\vs{{\sl vs.}}                      %vs. 
\def\sinthw{sin$^2 \theta_W$}           %sin^2 th_W
\def\mt{$m_t$}                          %m_top 
\def\mb{$m_b$}                          %m_bottom
\def\mw{$M_W$}                          %M_W
\def\mz{$M_Z$}                          %M_Z
\def\pizero{$\pi^0$}                    %pizero
\def\jpsi{$J/\psi$}                     %J/psi
\def\wino{$\widetilde W$}               %Wino
\def\zino{$\widetilde Z$}               %Zino
\def\squark{$\widetilde q$}             %squark
\def\gluino{$\widetilde g$}             %gluino
\def\alphas{$\alpha_{\scriptscriptstyle S}$}                %alpha_s
\def\alphaem{$\alpha_{\scriptscriptstyle{\rm EM}}$}         %alpha_EM
\def\epm{$e^+e^-$}                      %e+e-
\def\deg{$^\circ$}                      %degree sign
\def\met{\mbox{${\hbox{$E$\kern-0.6em\lower-.1ex\hbox{/}}}_T$ }} %missing ET
\def\mht{\mbox{${\hbox{$H$\kern-0.725em\lower-.1ex\hbox{/}}}_T$}} %missing HT
\def\st{\mbox{$S_T$}}          % same as \mht
%%
                        %% \def\draft{\special{header=draft.ps}}
%%      
%%  References help:  DO NOT USE THESE FOR PAPERS WHICH WILL BE SUBMITTED
%%                               TO PRL OR PRD!!!!!!!!!!
%%
\newcommand{\NC}{{\em Nuovo Cimento\/} }
\newcommand{\NIM}{{\em Nucl. Instr. Meth.} }
\newcommand{\NP}{{\em Nucl. Phys.} }
\newcommand{\PL}{{\em Phys. Lett.} }
\newcommand{\PR}{{\em Phys. Rev.} }
\newcommand{\PRL}{{\em Phys. Rev. Lett.} }
\newcommand{\RMP}{{\em Rev. Mod. Phys.} }
\newcommand{\ZP}{{\em Zeit. Phys.} }
% Parameters for references:
% parameter 1 - volume, parameter 2 - page, parameter 3- last two digits of year
\def\err#1#2#3 {{\it Erratum} {\bf#1},{\ #2} (19#3)}
\def\ib#1#2#3 {{\it ibid.} {\bf#1},{\ #2} (19#3)}
\def\nc#1#2#3 {Nuovo Cim. {\bf#1} ,#2(19#3)}
\def\nim#1#2#3 {Nucl. Instr. Meth. {\bf#1},{\ #2} (19#3)}
\def\np#1#2#3 {Nucl. Phys. {\bf#1},{\ #2} (19#3)}
\def\pl#1#2#3 {Phys. Lett. {\bf#1},{\ #2} (19#3)}
\def\prev#1#2#3 {Phys. Rev. {\bf#1},{\ #2} (19#3)}
\def\prl#1#2#3 {Phys. Rev. Lett. {\bf#1},{\ #2} (19#3)}
\def\rmp#1#2#3 {Rev. Mod. Phys. {\bf#1},{\ #2} (19#3)}
\def\zp#1#2#3 {Zeit. Phys. {\bf#1},{\ #2} (19#3)}

\hyphenation{PYTHIA}
\lefthyphenmin=3
\righthyphenmin=3
%\draft
\twocolumn

\newcommand{\mco}{\multicolumn}

%{\rightline {\small {\bf March 20, 2000}}}
\vskip 10pt % one lines
%\vspace*{0.10in}
\begin{center}
\begin{Large}   
{\bf Searches for Compositeness at the Tevatron\\}
\end{Large}

\begin{small}
{\bf ~\\J.~Andrew~Green}\\
Iowa State University\\
Ames, Iowa 50010, USA\\
agreen@fnal.gov\\
\vspace{\baselineskip}
on behalf of Fermilab. \\
Presented at Rencontres de Blois June 1, 1999,
with more recent searches now included.
\end{small}
\end{center}

\vskip 1.0cm
\normalsize
%\vspace{0.5cm}
%\bigskip

\subsection{Introduction}

The existence of three families of quarks and leptons suggests the 
possibility of a substructure for these objects \cite{ehlq}.  
The hypothetical constituents known generically as preons, 
interact via a new strong interaction called Metacolor.  
The characteristic energy scale, $\Lambda$, 
for the new interactions is, of course, unknown.  
The strength of the interactions 
through a contact term can be written as 
%\begin{equation}
$\hat{s}/(\alpha_S \Lambda^2 )$, 
%\end{equation}
where $\hat{s}$ is the square of the energy in the center
of mass frame of the (normal) interacting partons, 
and $\alpha_S$ is the QCD coupling. 

The first limit on the size of the atomic nucleus was obtained 
by Geiger and Mardsen in the Rutherford scattering of $\alpha$ particles 
from nuclei.  In an analogous way, we can set a limit on the 
size of quarks and leptons by observing the scattering of the highest energy 
quarks and antiquarks at the Fermilab Tevatron at 
$\overline{p}p $ center-of-mass energies of 1.8 TeV for collider experiments,
and 0.8 TeV for fixed-target experiments.  The collider detectors 
at Fermilab, CDF and \D0, have performed searches for  compositeness, 
and this paper gives a summary of those searches.  Those detectors 
are general-purpose, have nearly $4\pi$ acceptance, and measure lepton 
and jet energies to high precision.  In addition, the neutrino 
detector, CCFR, which utlilized the 800~\gev\ proton line at Fermilab 
has performed a compositeness search.

\subsection{Contact Interactions:  Indirect Searches for Compositeness}

Several previous searches for compositeness relied on direct
searches for excited quark
states \cite{excited_q}.  The searches that we discuss 
assume that $\sqrt{\hat{s}}$ is small compared to the characteristic 
mass scale, $\Lambda$.  These are generically called ``contact
interaction'' searches, and each interaction may be regarded 
as a contact term in the Lagrangian, which has the form

%\begin{equation}
%\begin{split}
%L = \frac{2\pi}{\Lambda^2}\Biggl\{
%&\eta^0_{LL}\lgroup\overline{q}_L\gamma^{\mu}q_L\rgroup\lgroup\overline{\psi}_L\gamma_{\mu}\psi_L\rgroup +
%\eta^0_{LR}\lgroup\overline{q}_L\gamma^{\mu}q_L\rgroup\lgroup\overline{\psi}_R\gamma_{\mu}\psi_R\rgroup +\\
%&\eta^0_{RL}\lgroup\overline{q}_R\gamma^{\mu}q_R\rgroup\lgroup\overline{\psi}_L\gamma_{\mu}\psi_L\rgroup +
%\eta^0_{RR}\lgroup\overline{q}_R\gamma^{\mu}q_R\rgroup\lgroup\overline{\psi}_R\gamma_{\mu}\psi_R\rgroup +\\
%&\eta^1_{LL}\lgroup\overline{q}_L\gamma^{\mu}{\frac{\Lambda_a}{2}}q_L\rgroup
%\lgroup\overline{\psi}_L\gamma_{\mu}{\frac{\Lambda_a}{2}}\psi_L\rgroup + \\
%&\eta^1_{LR}\lgroup\overline{q}_L\gamma^{\mu}{\frac{\lambda_a}{2}}q_L\rgroup
%\lgroup\overline{\psi}_R\gamma_{\mu}{\frac{\lambda_a}{2}}\psi_R\rgroup + \\
%&\eta^1_{RL}\lgroup\overline{q}_R\gamma^{\mu}{\frac{\lambda_a}{2}}q_R\rgroup
%\lgroup\overline{\psi}_L\gamma_{\mu}{\frac{\lambda_a}{2}}\psi_L\rgroup + \\
%&\eta^1_{RR}\lgroup\overline{q}_R\gamma^{\mu}{\frac{\lambda_a}{2}}q_R\rgroup
%\lgroup\overline{\psi}_R\gamma_{\mu}{\frac{\lambda_a}{2}}\psi_R\rgroup
%\Biggl\}
%\end{split}
%\end{equation}
%Where $\eta^0_{XX}$ is the singlet coefficient, and
%$\eta^1_{XX}$ is the octet coefficient. 
%$\eta^X_{LL}=\eta^X_{RR}=\eta^X_{LR}=\eta^X_{RL}=\pm 1$ corresponds to the vector
%interaction, and
%$\eta^X_{LL}=\eta^X_{RR}=-\eta^X_{LR}=-\eta^X_{RL}=\pm 1$ corresponds to the
%axial-vector interaction.  

\begin{equation}
\begin{split}
L = \frac{2\pi}{\Lambda^2}\Biggl\{
&\eta_{LL}\lgroup\overline{f}_L\gamma^{\mu}f_L\rgroup\lgroup\overline{f}_L\gamma_{\mu}f_L\rgroup + \\
&\eta_{LR}\lgroup\overline{f}_L\gamma^{\mu}f_L\rgroup\lgroup\overline{f}_R\gamma_{\mu}f_R\rgroup +\\
&2\eta_{RL}\lgroup\overline{f}_R\gamma^{\mu}f_R\rgroup\lgroup\overline{f}_L\gamma_{\mu}f_L\rgroup 
\Biggl\},
\end{split}
\end{equation}
where $f_{L,R}$ are the left and right-handed chiral components of the
the quark or lepton, and $\eta_{ij}$  is the sign of each
term, where $\eta_{ij}=+1$ corresponds to destructive, and 
$\eta_{ij}=-1$ to constructive interference \cite{ehlq}.

Only leading-order calculations are available for  
compositeness.  Hence, the following approximate relation 
is used to calculate the correction for higher-order
contributions:

\begin{equation}
\frac{\sigma(\Lambda=X\ )_{LO}}{\sigma(\Lambda=\infty\ )_{LO}}
\times\sigma(\Lambda=\infty\ )_{NLO or NLL},
\end{equation}
where $\Lambda=\infty\ $, corresponds to the standard model.

The next-to-leading order (NLO) or next-to-leading log (NLL) 
cross-sections are calculated using the
event generators {\sc jetrad} \cite{jetrad} or 
the Monte Carlo written by Hamburg {\it et al}\cite{d0_theory,cdf_theory}, 
respectively.  The presense of any compositness is
expected to contribute to
an increase in cross section at large
jet and lepton transverse momenta.  

Discussed in this paper are the most up-to-date
quark-quark and quark-lepton
contact interaction searches.  Of course, excellent lepton-lepton
searches have been accomplished as well \cite{opal}.

\subsection{Quark Compositenss} 

\subsubsection{Limits from Dijet Mass Measurements }

\D0 has measured \cite{dzero_dijet_mass} the ratio of the dijet mass spectrum 
for jets at pseudorapidities
$|\eta^{jet}|<0.5$ relative to jets at
$0.5<|\eta^{jet}|<1.0$.  
The resulting cancellation of systematic uncertainties results in 
an absolute systematic uncertainty of $< 8\%$ of the ratio.

\begin{figure}[hbt]
\begin{center}
%\begin{picture}(100,100)(-10,-5)
\mbox{\epsfxsize7.0cm\epsffile{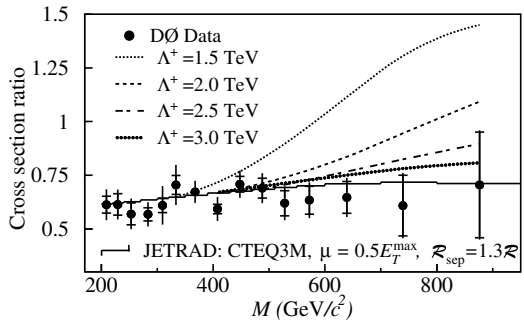}}
%\end{picture}
\caption{
The ratio of cross sections for jets with $|\eta_{jet}|<0.5$ to jets
with $0.5<|\eta_{jet}|<1.0$ for data (solid circles)
and theory (different $\Lambda$).  
The full error bars show the statistical
and systematic uncertainties added
in quadrature, and the crossbars show the size of the 
statistical uncertainties.
}
\label{d0_dijet_mass}
%\end{picture}
\end{center}
\end{figure}

\begin{figure}[hbt]
\begin{center}
%\begin{picture}(100,100)(-10,-5)
\mbox{\epsfxsize7.0cm\epsffile{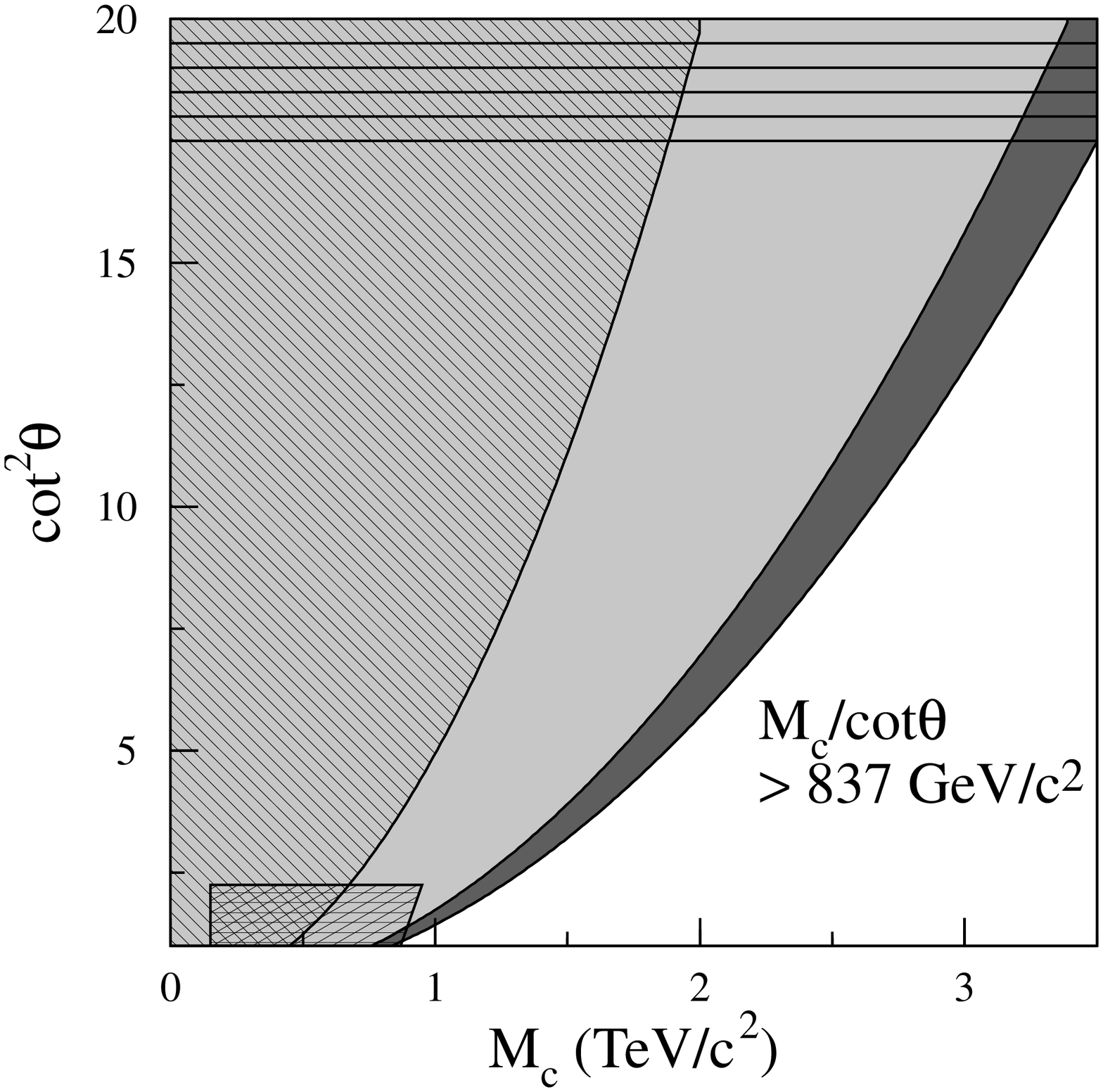}}
%\end{picture}
\caption{
Limits for the coloron parameter space: coloron mass $M_c$
vs. mixing parameter $\cot{\theta}$. The dark shaded region shows
the 95$\%$ CL exclusion region from the \D0 measurement of the ratio
of cross sections in Fig. \ref{d0_dijet_mass} 
(\mbox{$M_{c} / \cot{\theta} > 837 \gevcc$}). 
The lightly shaded region shows the area excluded by
the \D0 dijet angular distribution 
(\mbox{$M_{c} / \cot{\theta}> 759 \gevcc$}).
The horizontally hatched region at large
$\cot{\theta}$ is not allowed in this version of the model. The
diagonally hatched region is excluded by the value of $\rho$
(\mbox{$M_{c} / \cot{\theta} > 450 \gevcc$}). The cross--hatched
region at low $\cot^2\theta$ is excluded by the CDF search for new
particles decaying to dijets.
}
\label{d0_coloron}
\end{center}
\end{figure}

Predictions for quark compositeness are obtained using the LO 
{\sc pythia} event generator \cite{pythia}.  
The ratios of these LO predictions 
with, and without compositeness, are used to scale the {\sc jetrad} 
NLO predictions (see Eq. 2), 
and are compared to data in Fig.\ref{d0_dijet_mass}.  Limits on models
with color singlet (octet), vector, or axial contact
interactions were set using an analytic LO calculation \cite{chivukula}
rather than {\sc pythia}.  The resulting limits are given in Table
\ref{compositeness_limits_2}.  Models of quark 
compositeness with a contact interaction scale 2.4 TeV
for quark-quark interactions
are excluded at the 95\% CL.  The limits on the scale of 
$\Lambda^-_{V_8}$ can be converted into limits
on a flavor-universal coloron \cite{iain_coloron} resulting in a 95\%
CL of $M_C/cot\theta > 837 \gevcc$, where $M_C$ is the
mass of the coloron and $cot\theta$ a
parameter of the theory.  The comparison is given in Fig. \ref{d0_coloron}.

\setlength{\unitlength}{0.7mm}

\begin{table}[ht]
\begin{center}
\begin{minipage}[t]{2.4in}
\caption{95$\%$ confidence limits in TeV for different
 compositeness scale for different models calculated using an analytic
 LO prediction~{\protect\cite{chivukula}} (see text). }
 \label{compositeness_limits_2} 
 \begin{tabular}{ccc}
%\hline
 Model & $\displaystyle{\Lambda^{+}}$ & $\displaystyle{\Lambda^{-}}$ \\
\hline
 $\displaystyle{LL}$ (left-left isoscalar)   & 2.7 & 2.4 \\
 $\displaystyle{V}$ (vector singlet)    & 3.2 & 3.1 \\
 $\displaystyle{A}$ (axial singlet)   & 3.2 & 3.1 \\
 $\displaystyle{V_8}$ (vector octet) & 2.0 & 2.3 \\
 $\displaystyle{A_8}$ (axial octed) & 2.1 & 2.1 \\
%\hline
 \end{tabular}
\end{minipage}
\end{center}
\end{table}

\setlength{\unitlength}{0.7mm}

\subsubsection{Limits from The Dijet Angular Distribution}

\begin{figure}[htb]
\begin{center}
%\begin{minipage}[t]{2.45in}
{\epsfxsize=2.45in \epsfbox{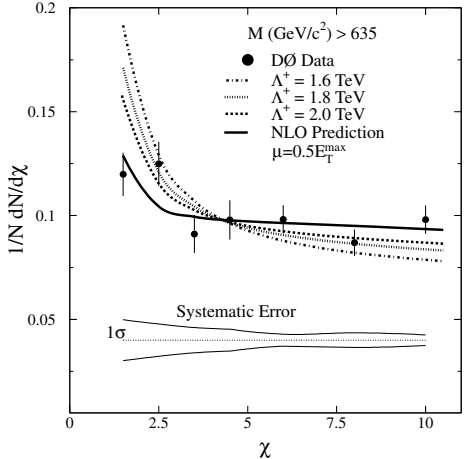}
\caption{\D0 dijet angular distribution compared to theory for
   different quark compositeness scales.The errors on the points
   are statistical and the band represents the correlated $\pm$ 1
   standard deviation systematic uncertainty.  } \label{fig_3} }
%\end{minipage}
\end{center}
\end{figure}

 At small center of mass scattering angles, $\theta^{\star}$,
 the dijet angular
 distribution predicted by leading order QCD is proportional to the
 Rutherford cross section: ${ {d\hat{\sigma}_{\rm ij} / {d\cos
 \theta^{\star}}} \sim {{1} / {\sin ^4({{\theta ^{\star}}/{2}})}}
 }$. It is useful to measure the angular distribution in the variable
 $\chi$, rather than $cos\theta^{\star}$, where $ {{\chi} = ({{1+\cos
 \theta^{\star}}) / ({1-\cos \theta^{\star}}}}) = e^{|\eta_1-\eta_2|}$.
 The variable $\chi$ facilitates comparison of
 theory with data. The angular distribution from 
 composite quarks in quark-quark interactions
 is isotropic.  Hence,
 the measurement of the dijet angular distribution is sensitive to the
 presence of such new physics.  The quantity measured in the dijet angular
 analysis is $\displaystyle{{1 / {N}} ({ {dN} / {d \chi}}) }$, which
 is measured in bins of dijet mass \jjmass. Figure~\ref{fig_3}
 depicts the angular distribution as measured\cite{d0_angular} by \D0
 for $ M_{\rm JJ}  > 625 \gevcc$.

\begin{figure}[htb]
\begin{center}
%\begin{minipage}[t]{2.45in}
{\epsfxsize=2.45in
\epsfbox{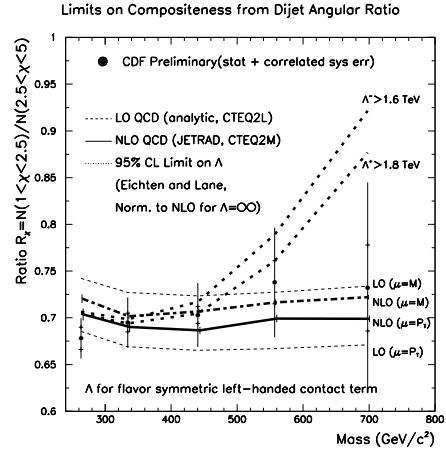}
\caption{CDF measurement of $R_{\rm \chi}$ as a function of dijet
   invariant mass.}
\label{fig_4}}
%\end{minipage}
\end{center}
\end{figure}

 To remove the point to point correlated errors, the distribution can
 be described by a single parameter,  $R_{\chi}$=N(${\chi < X })/N({ X <
 \chi < \chi_{\rm max} }$), which is the ratio of the number of events with
 $\chi < X$ to the number between $X < \chi <
 \chi_{max}$. The CDF\cite{cdf_angular} analysis uses $X = 2.5$, and
 D\O \cite{d0_angular} uses $X=4$. Figure~\ref{fig_4} exhibits $R_{\rm
 \chi}$ measured by CDF as a function of $M$ for two different
 renormalization scales, along with the predictions for
 different compositeness scales. The dijet angular
 distribution from QCD was calculated using the {\sc jetrad} event
 generator.  The predictions for quark compositeness scale
 are obtained using a LO analytic calculation\cite{ehlq}.  The ratio
 of these LO predictions with compositeness, to the LO with no
 compositeness, is used again to scale the {\sc jetrad} NLO predictions.

 Analysis of the CDF data exludes models with $\Lambda_{LL}^{+} 
 <$ 2.1 TeV and
 $\Lambda_{LL}^{-} <$ 2.2 TeV. In addition, \D0 
 uses their measurements to place limits
 on the production of colorons, requiring $M_{c} / \cot{\theta} >\ 
 759 \gevcc$ (see Fig. \ref{d0_coloron}).

\subsubsection{Limits from Large $H_T$}

\D0 has also used the $H_T = \sum{E_T^{jets}}$ 
variable to set limits on quark 
compositeness based on \cite{d0_large_ht}.  
By treating the event as a whole, 
this analysis complements 
the other searches for compositeness.  It studies
the compositeness of left-handed quarks in the 
left-left isoscalar contact term of the Lagrangian given in \cite{ehlq}.
The scale parameter for this term is $\Lambda_{\it LL}$.
 
The events are chosen to have  $H_T > 500~ \gev$,
well above the contribution expected from top quarks that peaks near  
$2 m_t \approx 350 \gev$.
Jets with $E_T > 20 \gev$ and
$|\eta^{jet}| < 3.0$ are included in the calculation of
$H_T$ for each event.  In addition, the events are required to
have at least one jet with $E_T > 115 \gev$.  Other cuts are 
applied to reduce the instrumental backgrounds, e.g.
from mis-vertexing.  The only important backgrounds considered are from 
such instrumental sources.  

Figure \ref{dth_figure} displays 
the fractional deviation between the data and the Monte Carlo
for the CTEQ4M PDF \cite{cteq} with renormalization scale 
$\mu=E_T^{\rm max}/2$.  
For $\Lambda_{\it LL}$ values between 
$1.4$ and $7.0 \tev$, {\sc pythia}\cite{pythia} is used to 
simulate the effects of quark 
compositeness to leading order.  This leading
order calculation is scaled using Eq. 2, where
{\sc jetrad} is used to compute the NLO component.
The results for composite quarks relative to 
expectations from the standard model are shown
in Fig. \ref{dth_figure} for $\Lambda_{\it LL} = 1.7, 2.0$ and $2.5 \tev$.
The ratios are found to be independent 
of the {\sc pythia} renormalization scale.

As seen in Fig. \ref{dth_figure}, quark compositeness would produce 
a rise in the cross section as a function of \htran.  
Changes in renormalization scale affect the absolute cross section, but 
not the shape of the \htran\ distribution.
Cross sections calculated using CTEQ4M or MRST \cite{mrst} PDFs 
differ in normalization but only slightly in shape. 

The measured \htran\ distribution above $500 \gev$ is well 
modeled by the {\sc jetrad} (NLO QCD) event generator. 
No evidence is found for compositeness in quarks, and 
the data are used to set limits on compositeness. As seen in 
Table \ref{limits-check-table}, the limits are not greatly
affected by choice of PDF.  The depencence of
the 95\% CL limit is also studied as a function of
renormalization scale, and shows little effect on the limits.
The limits for $\Lambda_{\it LL}^-$ are not given, but are
nearly identical (slightly higher than on $\Lambda_{\it LL}^+$).

\setlength{\unitlength}{0.7mm}
\begin{figure}[hbt]
\begin{center}
\begin{picture}(100,100)(-10,-5)
\mbox{\epsfxsize7.0cm\epsffile{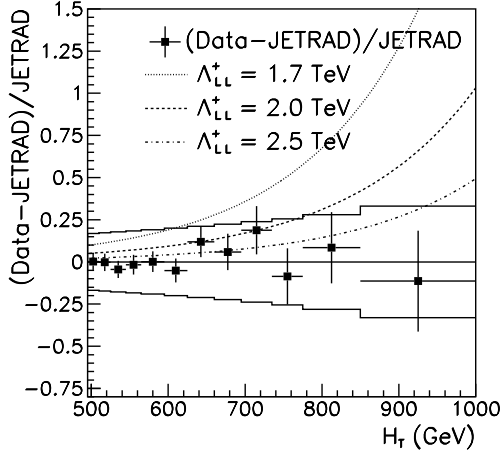}}
\end{picture}
\caption{Comparison of the measured \htran\ distribution with 
{\sc jetrad} (CTEQ4M and a
renormalization scale of $\mu = E_T^{\rm max}/2$).  
The errors on the points are statistical, and the 
error band represents the highly-correlated 
systematic uncertainty due to the jet energy 
scale.  The superimposed 
curves correspond to expectations for three compositeness scales.}
\label{dth_figure}
\end{center}
\end{figure}

\begin{table}[ht]
\begin{center}
\caption{The 95\% CL lower limits on quark compositeness 
scale in $\Lambda_{\it LL} (\tev)$
for different $\alpha_s$ (CTEQ4A1-5) and different gluon content 
(MRSTGU and MRSTGD).  
The renormalization scale is $E_T^{\rm max}/2$.}
\label{limits-check-table}
%\vspace{0.25cm}
\begin{tabular}{c cc cc c}
PDF&$\Lambda_{\it LL}^+$&PDF&
$\Lambda_{\it LL}^+$&PDF&$\Lambda_{\it LL}^+$\\ \hline
CTEQ4A1&2.0&CTEQ4A2&2.0&CTEQ4M&1.9 \\
CTEQ4A4&1.9&CTEQ4A5&1.9&& \\
MRSTGU&2.0&MRSTGD&2.1&MRST&2.0 
\end{tabular}
\end{center}
\end{table}

\subsection{Quark-Lepton Compositeness}

\subsubsection{Neutrino-Nucleon Scattering}

The CCFR experiment used the Fermilab 800 \gev proton beam directed at a 
Beryllium-Oxide target to produce a beam of neutrinos 
(86\% $\nu_{\mu}$, 12\% $\overline{\nu}_{\mu}$, 
and 2\% $\nu_e$($\overline{\nu}_e$)) with a mean energy
of 125 \gev.  The $\nu$'s are produced from decays of secondaries 
downstream from the BeO target.  The CCFR detector was a 
$3 \times 3 \times 18m$ neutrino sampling calorimeter with the ability to 
resolve minimum-ionizing tracks (mostly from muons) and determine 
shower energy.  The CCFR experiment ran from 1984 to
1988 and $\sim 1\times 10^6 \ \nu$ scatterings 
pass typical analysis cuts.
This detector has evolved to be what is now known as NuTeV,
which used an experimental program of switching between 
$\nu$ and $\overline{\nu}$ beams to drastically 
reduce systematic errors and improve 
measurements of quantities such as $sin^2 \theta_W$, and $M_W$
\cite{nutev}.

The primary goal of these detectors was to do high-statistics studies
of charged and neutral current interactions via $\nu N$ 
scattering.  Most of these studies rely on measuring the ratio of 
the Neutral Current (NC) to Charged Current (CC) interactions seen in the
detector.  
The NC interactions produce a compact shower of energy, while the CC
interactions produce a shower, with a long ($> 2 m$) 
minimum-ionizing trail left by the \mbox{$\mu$ or $\overline{\mu}$}.
An important correction to the ratio comes from the 
$\nu_e$($\overline{\nu}_e$) flux component, 
where the CC interaction produces 
$e$'s that, unlike $\mu$'s, cannot as easily be distinguished 
from the hadronic shower in a full range of energies.  Such events
can fake the NC signature.

The $R_{meas}=NC/CC$ ratio is used to measure the coupling, $\kappa$, of the 
$Z^0$ boson to quarks, where the error in the comparison of $R_{meas}$ to 
theory is about 0.6\%.
%Principal uncertainties in the theoretical comparison of $R_{meas}$ to data,
%are the uncertainty in the $\nu_e$ flux, and statistical error in the data.
With this measurement, CCFR has set limits on 
quark-neutrino contact interactions \cite{ccfr}.   
Table \ref{ccfr_limits} shows these limits.  

The NuTeV collaboration will soon provide complimentary and 
more precise information 
which can further constrain these contact interactions \cite{nutev}.

\begin{table}[ht]
\begin{center}
\caption{95\% CL lower limits on mass scales of new
$\nu \nu qq$ contact terms from CCFR.}
\label{ccfr_limits}
\begin{tabular}{c c c}
Interaction&$\Lambda^+$ (TeV) &$\Lambda^-$ (TeV)\\ \hline
LL&4.7&5.1 \\
LR&4.2&4.4 \\
RL&1.3&1.8 \\
RR&3.9&5.2 \\
VV&8.0&8.3 \\
AA&3.7&5.9 \\
\end{tabular}
\end{center}
\end{table}

\subsubsection{Drell-Yan Production}

 Both CDF\cite{cdf_drell_yan} and \D0 \cite{d0_drell_yan} have placed
 limits on the combined quark-electron 
 compositeness\cite{qe_compos} scale by
 analyzing Drell-Yan dilepton production.

\begin{figure}[htb]
\begin{center}
%\begin{minipage}[t]{2.45in}
{\epsfxsize=2.45in
\epsfbox{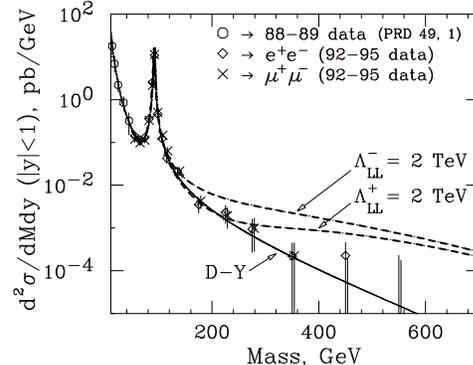}
\caption{The Drell-Yan cross section as measured by CDF for muons (crosses) and
 electrons (diamonds). The data are normalized between 50
 and 150 \gev/c$^2$ to the standard model cross-section. The curves are
 based on 
 standard model NLL calculation and $\Lambda$.} 
\label{fig_6} }
%\end{minipage}
%\hspace{2mm}
%\begin{minipage}[t]{2.45in}
\end{center}
\end{figure}

 CDF has measured the Drell-Yan cross section for both electrons and
 muons, satisfying $|\eta| < 1.0$ for each lepton 
 (see Fig.~\ref{fig_6}). The backgrounds to the signal from
 QCD jets mimicking electrons, from $W^{+}W^-$, $\tau^{+}\tau^-$, and heavy
 quark production, are estimated using data, and subsequently
 subtracted. The Drell-Yan production cross section is based on
 an NLL calculation\cite{cdf_theory}, 
 with and the data normalized to the
 prediction in the mass range between 50 and 150 {\rm GeV}/c$^2$. 
 The data are then
 used to place limits on the combined quark-electron compositeness (see
 Table~\ref{table_2}).

\begin{figure}[htb]
\begin{center}
{\epsfxsize=2.43in
\epsfbox{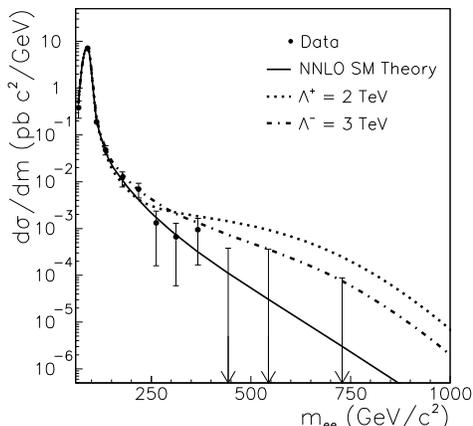}
\caption{The electron-electron cross section as measured by \D0,
and expected for Drell-Yan and Drell-Yan including
contact interactions (both including backgrounds).  Error bars
indicate statistical errors. }
\label{fig_7}}
%\end{minipage}
\end{center}
\end{figure}

\begin{table}[htb]
\begin{center}
\caption{The 95\% CL lower limit on the combined quark-lepton compositeness
scale $\Lambda$ for different \mbox{contact-interaction} models.}
\label{table_2}
\begin{tabular}{c|cc|cc|cc||cc}
%\hline
 & \multicolumn{6}{c||}{CDF} &  \multicolumn{2}{c}{\D0 } \\
\hline
Model & $\displaystyle{\Lambda_{qe}^{+}}$ & $\displaystyle{\Lambda_{qe}^{-}}$ 
& $\displaystyle{\Lambda_{q\mu}^{+}}$ & $\displaystyle{\Lambda_{q\mu}^{-}}$    
& $\displaystyle{\Lambda_{ql}^{+}}$ & $\displaystyle{\Lambda_{ql}^{-}}$ 
& $\displaystyle{\Lambda_{qe}^{+}}$ & $\displaystyle{\Lambda_{qe}^{-}}$ \\
\hline
LL&	2.5&	3.7&	2.9&	4.2&	3.1&	4.3&	3.3&	4.2\\
LR&	2.8&	3.3&	3.1&	3.7&	3.3&	3.9&	3.4&	3.6\\
RL&	2.9&	3.2&	3.2&	3.5&	3.3&	3.7&	3.3&	3.7\\
RR&	2.6&	3.6&	2.9&	4.0&	3.0&	4.2&	3.3&	4.0\\
VV&	3.5&	5.2&	4.2&	6.0&	5.0&	6.3&	4.9&	6.1\\
AA&	3.8&	4.8&	4.2&	5.4&	4.5&	5.6&	4.7&	5.5\\
%\hline
\end{tabular}
\end{center}
\end{table}

 \D0 has also measured the Drell-Yan cross section for electrons
 satisfying either $|\eta| < 1.1$ or $1.5 < |\eta| < 2.5$  
 (Fig.~\ref{fig_7}). \D0 measures
 the contribution to the cross section from misidentified QCD
 events, and other processes, using a combination of data and {\sc
 pythia} Monte Carlo.  The production cross section for
 Drell-Yan is calculated using an NNLO
 prediction\cite{d0_theory}. Limits are placed on various models of
 quark-electron compositeness (Table~\ref{table_2}).

 \D0 and CDF rule out models of quark-electron compositeness with
 interaction scales below 2.5 to 6.3 TeV, depending on the details of
 the model. 
%These limits are the most stringent available on
% quark compositeness.

\subsection{Conclusion}

The Fermilab experiments, CDF, \D0, and CCFR have shown that 
predictions from the Standard Model are in excellent agreement with
the data, and there is no evidence for compositeness 
in quarks below a scales from 2-8 TeV.  

The next run of the Tevatron, beginning in early 2001,
will reach far greater luminosities, and provide more opportunity for 
finding new physics at higher mass scales.

\subsection{Acknowledgements}

The author would like to acknowledge assistance from Iain Bertram
in the preparation of the talk and this proceeding.  

%\newpage

\end{document}